\documentstyle[11pt,epsfig]{article}

\newcommand{\bee}   {\begin{equation}}
\newcommand{\ene}   {\end{equation}}
\newcommand{\beqa}  {\begin{eqnarray}}
\newcommand{\enqa}  {\end{eqnarray}}
\newcommand{\bea}   {\begin{array}}
\newcommand{\ena}   {\end{array}}
\newcommand{\dy} {\displaystyle}
\newcommand{\gapproxeq}{\lower .7ex\hbox{$\;\stackrel{\textstyle >}{\sim}\;$}}

\def\ru1{\rule[-0.4truecm]{0mm}{1truecm}}
\def\al{\alpha}
\def\be{\beta}
\def\de{\Delta}
\def\als{{\alpha_s}}
\def\upa{\uparrow}
\def\dwa{\downarrow}

\begin{document}
\thispagestyle{empty}

$~$
\hfill{ \mbox{{\bf Bari-TH/97-272}}}

\hfill{ \mbox{Napoli Preprint {\bf DSF-T-97/21}}}

\hfill{\mbox{hep-ph/9705475}}

\vspace{2truecm}

{\LARGE 
\centerline{A Positive Test}
\vspace{.3truecm}
\centerline{for Fermi-Dirac Distributions}
\vspace{.3truecm}
\centerline{of Quark-Partons}}

\vspace{2truecm}

{\large 
\centerline{Franco BUCCELLA$^{a,b}$, Ilya DOR\u{S}NER$^c$, Ofelia
PISANTI$^b$,} 
\vspace{.4truecm}
\centerline{Luigi ROSA$^a$, and Pietro SANTORELLI$^d$}}

\vspace{0.5truecm}

\begin{center}
{\footnotesize {\it
a) Dipartimento di Scienze Fisiche, Universit\`a ``Federico II'', \\
Pad. 19 Mostra d'Oltremare, 80125 Napoli, Italy \\
and INFN, Sezione di Napoli, \\
Pad. 20 Mostra d'Oltremare, 80125 Napoli, Italy \\[.3truecm]
b) Istituto di Fisica Teorica, Universit\`a ``Federico II'', \\
Pad. 19 Mostra d'Oltremare, 80125 Napoli, Italy \\[.3truecm]
c) Privodnu-Matemati\v{c}ki Fakultet, Univerzitet u Sarajevu, \\
Zmaja od Bosne 32, Sarajevu, Bosna i Hercegovina \\[.3truecm]
d) Dipartimento di Fisica dell'Universit\`a di Bari, \\
and INFN, Sezione di Bari, \\
via G.Amendola 173, 70126 Bari, Italy \\
}}
\end{center}

\vspace{.9truecm}

\begin{abstract}
By describing a large class of deep inelastic processes with standard
parameterization for the different parton species, we check the
characteristic relationship dictated by Pauli principle: broader shapes for
higher first moments. Indeed, the ratios between the second and the first
moment and the one between the third and the second moment for the valence
partons is an increasing function of the first moment and agrees
quantitatively with the values found with Fermi-Dirac distributions. 
\end{abstract}

\newpage
\baselineskip   = 18pt

Four experimental facts conspire to indicate that the Pauli exclusion
principle plays a role in the quark-parton distributions in the nucleons.
They are the defect \cite{nmc} in the Gottfried sum rule \cite{gottfr}, the
asymmetry in the Drell-Yan processes on proton and deuteron targets
\cite{na51}, and the high $x$ behaviour of the ratios $F_2^n(x)/F_2^p(x)$
\cite{higx} and $g_1^p(x)/F_1^p(x)$ \cite{emc}. The first two facts imply
the inequality $\bar d > \bar u$ in the sea of the proton, which has been
advocated long time ago \cite{fife}. From the other two one can deduce the
dominance of the $u^{\upa}$ parton in the high $x$ region; this is an
indication of the relationship between the shape of the distribution as a
function of $x$ and the first moment of a parton required by Pauli
principle: broader shapes for higher first moments. 

More recently, by assuming the approximate relationship $u^{\dwa}(x) =
\dy\frac{1}{2}d(x)$, it has been possible to successfully relate
\cite{buso} the structure functions $F_2^p(x) -F_2^n(x) $ and $xg_1^p(x)$
for $x \geq 0.2$. 

Pauli principle suggests to assume \cite{bobu} Fermi-Dirac functions for
the quark distributions, 
\bee
q^{\upa(\dwa)}(x) = \frac{f(x)}{
                 \exp\{\frac{x-\tilde{x}(q^{\upa(\dwa)})}{\bar x}\} + 1 }\,,
\label{e:FD}
\ene
and Bose-Einstein functions for the gluons,
\bee
G^{\upa(\dwa)}(x) = \frac{8}{3}\frac{f(x)}{
                 \exp\{\frac{x-\tilde{x}(G^{\upa(\dwa)})}{\bar x}\} - 1 
}\,,
\label{e:BE}
\ene
where $\bar x$ plays the role of the {\it temperature}, the $\tilde x$ are
the {\it thermodynamical potentials} of each parton species and $f(x)$ is a
{\it weight function} with the usual form $A~ x^\al~ (1-x)^\be$. 

To account for the increase of $F_2^p(x)$ at small $x$ \cite{hera} we have
to add to Eq.~(\ref{e:FD}) an additional contribution $q_L(x) = \bar
q_L(x)$, which should be unpolarized and isospin-invariant to get finite
quark-parton model sum rules (QPMSR). This leads to a satisfactory
description of a large class of deep inelastic data in Ref.~\cite{bmt} and
its updated version \cite{erice}, including the $E154$ data by SLAC
\cite{e154}. 

Here we want to test a specific property of the quantum statistical
distributions given by Eq.~(\ref{e:FD}), namely the shape-first moment
relationship previously mentioned for fermionic partons. The QPMSR imply 
\bee
u^{\upa} \gg d^{\dwa} \gapproxeq u^{\dwa} > d^{\upa}\,.
\label{e:abun}
\ene
The defect in the Gottfried sum rule yields
\bee
\bar d > \bar u\,.
\label{e:abunqb}
\ene
We can argue the same pattern than Eq.~(\ref{e:abun}) and (\ref{e:abunqb})
for the corresponding broadnesses. To be more precise we expect the ratio
between the second and the first moment, that is the mean value of $x$, as
well as the one between the third and the second moment of each quark
parton to be an increasing function of the first moment. For the gluons the
deviations from the Boltzmann distribution (implying the shape to be
independent from the first moment) are in the opposite direction; thus, one
expects $\de G(x)$ softer than $G(x)$, and $G(x)$ softer than $\bar q(x)$,
if the same $f(x)$ is taken for quarks and gluons. 

To establish whether experiments confirm this shape-first moment
relationship, we try to describe the same data considered in \cite{erice},
by taking, for each light quark-parton species of a given flavour and spin,
at $Q_0^2\,=\,3\,GeV^2$, the form ($p=u^\upa,~u^\dwa,~d^\upa,~d^\dwa,~\bar
u^\upa=\bar u^\dwa,~\bar d^\upa=\bar d^\dwa$): 
\bee
p(x,Q_0^2) = A_p x^\al (1-x)^{\be_p} + A_L x^{\al_L} (1-x)^{\be_L}
\,.
\label{e:distr}
\ene
Going on with the {\it thermodynamical} language, we name {\it gas} and
{\it liquid} the two terms in Eq.~(\ref{e:distr}). 

We are not very sensitive to the strange quark distribution. Indeed, the
only observables depending on it that we are considering are
$F_2^n(x)/F_2^p(x)$ and $F_2^p(x)$. For the first one isospin invariance
implies the same contribution for both members of the ratio, the latter
being near to one in the small $x$ region, where strange partons are
concentrated. In that region the second observable, $F_2^p(x)$, receives a
large contribution from the second term of Eq.~(\ref{e:distr}), responsible
for its increase at small $x$. Therefore, in order to avoid the
introduction of new parameters, we take, according to Ref.~\cite{ccfr}, 
\bee
s(x) = \bar s(x) = \frac{\bar u(x) + \bar d(x)}{4.2}\,.
\ene

As in Ref.~\cite{erice} we consider $F_3(x)$, measured at $Q^2\, =\,3\,
GeV^2$ \cite{xf3}, $F_2^p(x)$ and $F_2^n(x)$ at $Q^2\, =\,4\, GeV^2$,
\cite{nmc,hera,f2p} and the polarized structure functions measured at SLAC
($g_1^p(x)$ and $g_1^d(x)$ at $Q^2\, =\,3\, GeV^2$ \cite{e143}, $g_1^n(x)$
at $Q^2\, =\,2\, GeV^2$ \cite{e142} and $Q^2\, =\,5\, GeV^2$ \cite{e154}).
To report all the data at the same $Q_0^2\, =\,3\, GeV^2$ we assume the
$Q^2$ dependence found by NMC, while the good agreement, in the common $x$
range, between the data regarding $g_1^n(x)$ at $Q^2\, =\,2\, GeV^2$ and at
$Q^2\, =\,5\, GeV^2$ makes us confident that they both give a good
approximation of that quantity at $Q^2\, =\,3\, GeV^2$. 

The unpolarized structure functions are given by:
\beqa
F_2^{p,n} (x) &=& x~ \left\{ \pm~ \frac{1}{6}~ [u(x)+\bar u(x)-d(x)-\bar
d(x)] + \frac{5}{18}~ [u(x)+\bar u(x)+d(x)+\bar d(x)] \right. \nonumber \\ 
&& \left. +~ \frac{1}{9}~ [s(x)+\bar s(x)] \right\}\,, \\
x~F_3 (x) &=& x~ [u(x)+d(x)+s(x)-\bar u(x)-\bar d(x)-\bar s(x)]\,.
\nonumber 
\enqa
As long as for the polarized distribution we recall that the presence of a
gluon isoscalar contribution, related to the anomaly, has been advocated
\cite{anom} to explain the defect in the Ellis-Jaffe sum rule \cite{elljaf}
for the proton found by the EMC experiment \cite{emc}. This contribution,
up to first order in $\als$, is given by the convolution \cite{conv} 
\bee
S(x)= - \frac{\als}{6 \pi} \int_x^1 \frac{dz}{z}\, (1-2\,z) \left( ln
\frac{1-z}{z} - 1 \right)\, \de G \left(\frac{x}{z}\right)\,. 
\label{e:conv}
\ene
In the framework of a complete NLO analysis, we should consider all
possible $\als$ contributions, but this is beyond the purposes of the
present work. However, in a factorization scheme where chiral symmetry is
respected (a choice closer to the picture of the nucleons in term of
constituent quarks \cite{altla}), the high value of $\de G$ required to
explain the experimental results suggests that the gluon polarization gives
the dominant contribution proportional to $\als$. Therefore, we shall give
the following expressions for the nucleon polarized structure functions: 
\bee
g_1^{p,n} (x) = \pm~ \frac{1}{12}~ [\de u(x)-\de d(x)] + 
\frac{5}{36}~ [\de u(x)+\de d(x)] + S(x)\,, 
\ene
with
\bee
\de G(x,Q_0^2) = A_{\de G}\; x^{\al_{\de G}} \;
(1-x)^{\be_{\de G}}\,. 
\label{e:dg}
\ene
Eqs.~(\ref{e:conv}) and (\ref{e:dg}) give a three parameter description of
$S(x)$ ($\als~A_{\de G},~\al_{\de G}$, and $\be_{\de G}$), which is general
enough for the gluon isoscalar term that one has to add to the valence term
to obtain $g_1^p(x)$ and $g_1^n(x)$. However, the fact that the presence of
this contribution, which is important in the small $x$ region, influence
only a little the determination of the shapes of the valence quarks, which
depend on the $\be_p$'s, makes the debate on this issue purely academic
with respect to the main purpose of this paper. 

As usual, we assume
\bee
g_1^d (x) = \dy \frac{1}{2} \left( 1 - \frac{3}{2} \omega_D \right)
(g_1^p(x) + g_1^n(x))\,,
\ene
where $\omega_D$ is the amount of D-wave in the deuteron ground state
(0.058) \cite{omD}. 

We have then four parton distributions ($u$, $d$, $\bar d$ and $\bar u$) to
describe the three unpolarized structure functions ($F_2^{p}(x)$,
$F_2^{n}(x)$, and $F_3(x)$), and three parton distributions ($\de u$, $\de
d$ and $\de G$) for the two polarized ones ($g_1^p(x)$ and $g_1^n(x)$). To
overcome the ambiguity arising from this situation, we require the
unpolarized distributions to comply with the Adler sum rule \cite{adler}
and with the NA51 \cite{na51} measurement of the asymmetry in the Drell-Yan
production of muon pairs, 
\bee
\al_{DY}(x=.18) = -0.09 \pm 0.04 \pm 0.02\,.
\ene
We also require the $\be_{\bar q}$  and $\be_{\de G}$ to be larger than all
the $\be_q$'s and 
\bee
-0.8\; < \; \al,~\al_{\de G}\;\leq \; 0\;.
\ene
We consider the cases with $\de G=0$ and with $\de G$ free. At small $x$ it
is difficult to separate the two terms in Eq.~(\ref{e:distr}). Since the
second term is expected to give a small contribution ($\sim 10 \%$)
\cite{bmt} to the second moment of each parton, but an infinite one to the
first moment, we expect the second moment of the first term to be less
affected than the second one by the ambiguity in disentangling the two
contributions. However, the second term gives an equal contribution to each
quark-parton: therefore the existence of a correlation between the
contribution to the first moment and the shape of the first term in
Eq.~(\ref{e:distr}) is not affected by this uncertainty, which has the same
influence on the first and second moment of the {\it gas} component of each
parton. Moreover, since the low $x$ behaviour of the two contributions is
very different (in the case of Ref.~\cite{bmt} their power behaviours
differ by almost one unity), we expect that also for the first moment the
ambiguity is not so large. 

As we have previously said, the Fermi-Dirac function for the quark
distributions gives an automatic correlation between the first moment and
the shape of each quark-parton distribution, namely broader shapes for
higher first moments, with both quantities depending on $\tilde x$, once
fixed $f(x)$ and $\bar x$. Instead, for the first term of
Eq.~(\ref{e:distr}), once imposed a common value for $\al$, there are two
parameters for each parton, $A_p$ and $\be_p$, to reproduce the first two
moments of the first term in Eq.~(\ref{e:distr}), which we shall indicate
with $p$ and $p^{(2)}$ respectively. The second term in
Eq.~(\ref{e:distr}), responsible of the increase of $F_2^p (x)$ at low $x$
and apparently power-like, gives an infinite contribution to any $p$ (equal
for each light parton, should we start from a finite positive value of $x$)
and an equal contribution to $p^{(2)}$. 

Since it is difficult to disentangle the {\it gas} and {\it liquid}
component in Eq.~(\ref{e:distr}), we parameterize each light parton
distribution with the same $\al$; in this way the shape of the
corresponding distribution is dictated by $\be_p$, which is related to the
high $x$ behaviour, where the contribution of the {\it liquid} part is
expected to become negligible. To make the comparison between
Eqs.~(\ref{e:FD}) and (\ref{e:distr}) more appropriate, we fix the {\it
liquid} part to the value previously found with the parton statistical
distributions, namely 
\bee
A_L~ x^{\al_L}~ (1-x)^{\be_L} = 
\bea{ll}
0.12~ x^{-1.19}~ (1-x)^{9.8} & \de G = 0\,, \\
\\
0.12~ x^{-1.19}~ (1-x)^{9.6} & \de G~ \mbox{\rm free}\,.
\ena
\label{e:liq}
\ene

In tables \ref{t:param}, \ref{t:mom1a} and \ref{t:mom1b} the values of the
parameters found in the fit and the first three moments of the {\it gas}
component, with their ratio for each parton, with and without gluons, are
reported and compared with the updated fit (whose results are almost
identical to the one in Ref.~\cite{erice}) with quantum statistical
distributions.\\ 
An attentive glance to the numbers in the tables brings us to the following
conclusions: 
\begin{itemize}
\item[{i)}] for the quantum statistical as well as for the present
parameterization the parameters and the moments of the quark-partons have a
very weak dependence on the gluon contribution and the agreement between
the moments, obtained with the different parameterization, is good with the
exception of the $\bar u$ parton and $\de G$. 
\item[{ii)}] using parameterization (\ref{e:distr}) we get, for the ratios
$p^{(2)}/p$ and $p^{(3)}/p^{(2)}$ as a function of $p$ for the quarks, a
pattern very similar to the one obtained with Fermi-Dirac functions, which
have the property 
\beqa
\frac{u^{\upa (2)}}{u^{\upa}} &\gg& \frac{d^{\dwa (2)}}{d^{\dwa}}
\gapproxeq \frac{u^{\dwa (2)}}{u^{\dwa}} > \frac{d^{\upa (2)}}{d^{\upa}}\,, 
\\
\frac{u^{\upa (3)}}{u^{\upa (2)}} &\gg& \frac{d^{\dwa (3)}}{d^{\dwa (2)}}
\gapproxeq \frac{u^{\dwa (3)}}{u^{\dwa (2)}} > \frac{d^{\upa (3)}}{d^{\upa 
(2)}}\,; \nonumber
\nonumber 
\enqa
this shows that the ratios $p^{(2)}/p$ and $p^{(3)}/p^{(2)}$ are increasing
functions of $p$ according to Eq.~(\ref{e:abun}). Also, $\bar d^{(2)}/\bar
d > \bar u^{(2)}/\bar u$, as we expect from Eq.~(\ref{e:abunqb}). 
\end{itemize}

The differences found for $\bar u$ and $\de G$ are not unexpected. As for
the first one, which is the lighter parton with the lowest first moment,
the difference is large in percentage but small in amount. The different
shape found, which is a consequence of the high value of $\be_{\bar u}$, is
explained by the small precision in the determination of a parameter which
is related to the shape of a distribution with a very low first moment.
Concerning the gluons, we are not surprised to find a $\de G$ about more
than twice larger than with the Bose-Einstein form, which has a less
divergent extrapolation to small $x$, and a different shape: the fact that
the fit is equally good with and without gluons shows that we are not able
to get, within our approach, relevant information on them. However, the
value found is consistent with the NLO result of Ref.~\cite{newnlo}, $\de G
= 1.1 \pm 0.4$, where the gluons are also constrained by the $Q^2$
dependence of the data. The value found for $\al_{\de G}$ is at the lower
limit of the one in Ref.~\cite{newnlo}. 

The fact that the ratios $d^{\upa (2)}/d^{\upa}$ and $d^{\upa (3)}/d^{\upa
(2)}$, for the valence quark with the smallest first moment, are smaller
than the ones for the other valence quarks confirms the correlation among
broader shape and higher first moment, which shows up in the dominance of
$u^\upa$ at high $x$. Needless to say, $u^{\upa (2)}/u^{\upa}$ and $u^{\upa
(3)}/u^{\upa (2)}$ come out larger than the corresponding ratio of the
other valence partons, as we expect from the high $x$ behaviour of $F_2^n
(x)/F_2^p (x)$ and $A_1^p (x)$. Indeed, already in Ref.~\cite{nic}, by
taking the single parton distributions from a set of deep inelastic data,
the author has been able to find a similar trend with a distribution for
$u^\upa$ and for $d^\upa$ broader and narrower respectively than for the
other valence partons. 

The values of the first and second moment of the {\it gas} part in
Eqs.~(\ref{e:FD}) and (\ref{e:distr}) are in good agreement. (The exception
of $\bar u$ does not contradict this conclusion, since the small value of
its first moment makes less relevant its shape.) This agreement is, at our
advice, an additional argument in favour of the Fermi-Dirac distributions,
for which, once fixed $\bar x$ and $f(x)$ for each parton, $p$, $p^{(2)}$,
and $p^{(3)}$ depend on one parameter, $\tilde x (p)$; to say it in a more
clear way, the fit with the present parameterization might give equal
values of $\be_q$'s and distributions which differ only by a constant
factor, which would be, within our statistical language, the Boltzmann
limit. The agreement found for the ratio $p^{(3)}/p^{(2)}$ for the valence
quarks with the two different parameterizations gives a positive test,
which is very little affected by the difficulty of disentangling the gas
from the liquid contribution, since the last one has little relevance in
the evaluation of the second moments of the valence quarks and even a minor
one for the third moments. 

We wish to compare our evaluation of the first two moments of the
non-singlet polarized distributions with a previous analysis of the same
data \cite{bupi} and with the result found in the NLO analysis of
Ref.~\cite{newnlo}. To this purpose, in table \ref{t:mom2} these moments
and their ratio are reported for $g_1^{p(NS)}$. One can see a very good
agreement of the value found for the second moment of $g_1^{p(NS)}$ between
the quantum statistical distributions, the present parameterization and the
one in \cite{newnlo}, while the difference found for the first moment
depends on the small $x$ behaviour. 

The gluons are expected to have a softer distribution than the $\bar q$'s
if the same $f(x)$ is taken. The great uncertainty on their distribution
does not allow to draw meaningful conclusions from the fact that table
\ref{t:mom1b} agrees with this expectation. A typical property of the
Bose-Einstein function is that $\de G(x)/G(x)$ is a decreasing function of
$x$ (soft polarization for the gluons), therefore $\de G^{(2)}/\de G <
G^{(2)}/G$. In order to show this property one needs a NLO analysis
suitable to determine the shape of $\de G$ and $G$. We plan to perform a
consistent NLO analysis to include all the contributions proportional to
$\als (Q^2)$ and to evolve the distributions found in this way to the
higher $Q^2$ ($\sim 10\,GeV^2$) of the CERN experiments on polarized proton
and deuteron targets. We do not expect a too challenging test for the
resulting predictions, since the precision of the measurements, especially
in the low $x$ region not reached at SLAC, is worse for CERN data. 

In fact, we tried to extend our analysis to the CERN data at $<\!\!Q^2\!\!>
= 10\,GeV^2$ for $g_1^p (x)$ and $g_1^d (x)$ \cite{smc}, together with the
unpolarized distributions evaluated at $Q^2=10\,GeV^2$. Unfortunately the
absence of data on $g_1^n$ ($g_1^{H\!e^3}$) and the minor precision of the
data for $g_1^d$, together with the fact that the unpolarized structure
functions at $Q^2=10\,GeV^2$ are known with a smaller precision, do not
allow to extract from the data the trend of $p^{(2)}/p$ and
$p^{(3)}/p^{(2)}$, which come out different from equally good fits. 

\begin{figure}[t]
\epsfig{file=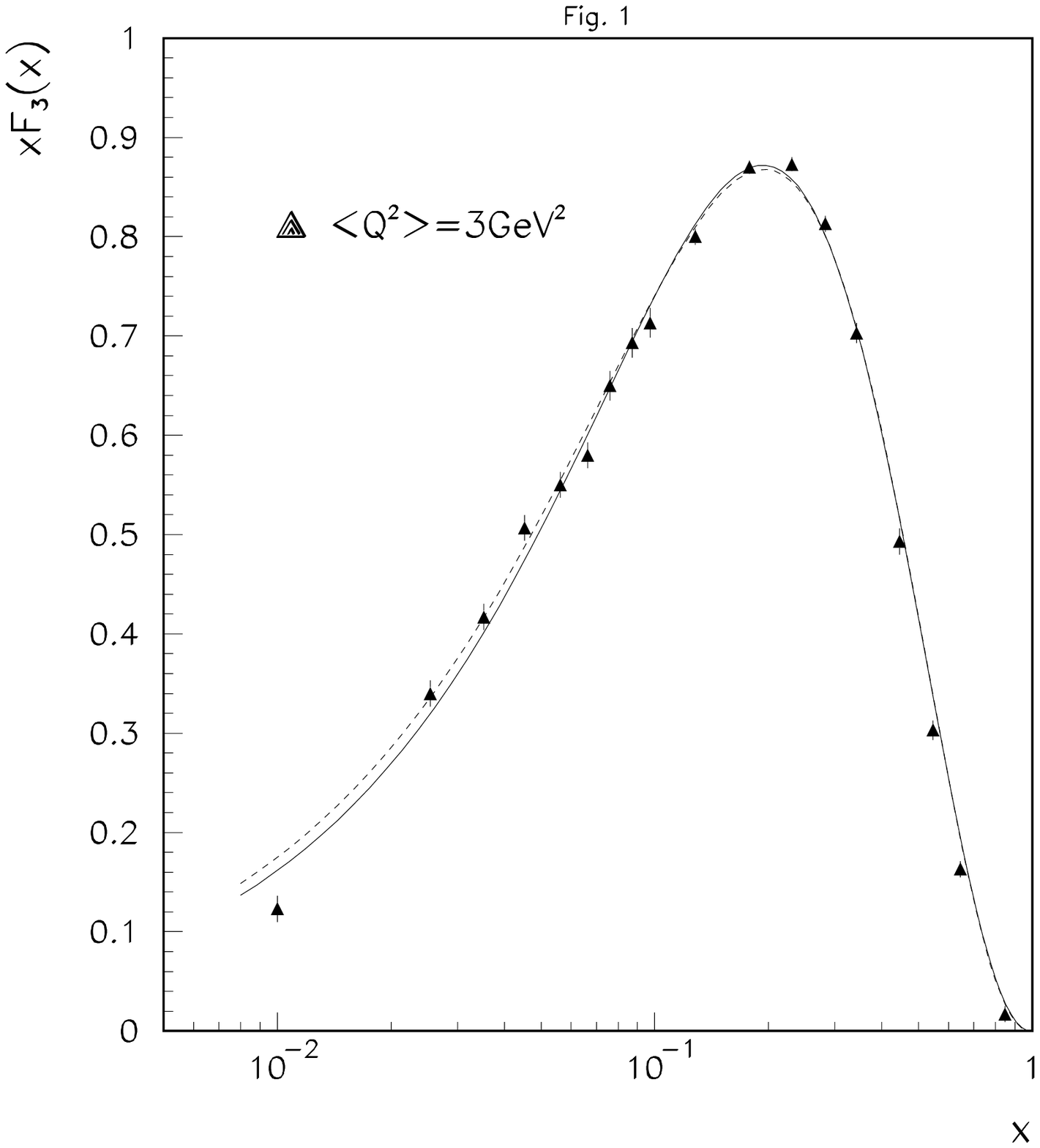,height=15cm}
\caption{The prediction for $x F_3(x)$ at $Q^2=3~GeV^2$ in the case $\de G$
free is plotted and compared with the experimental data \protect\cite{xf3}.
The solid and dashed line corresponds to the fit with present
parameterization and with quantum statistical distribution respectively.} 
\label{f:fig1}
\end{figure}

\begin{figure}[t]
\epsfig{file=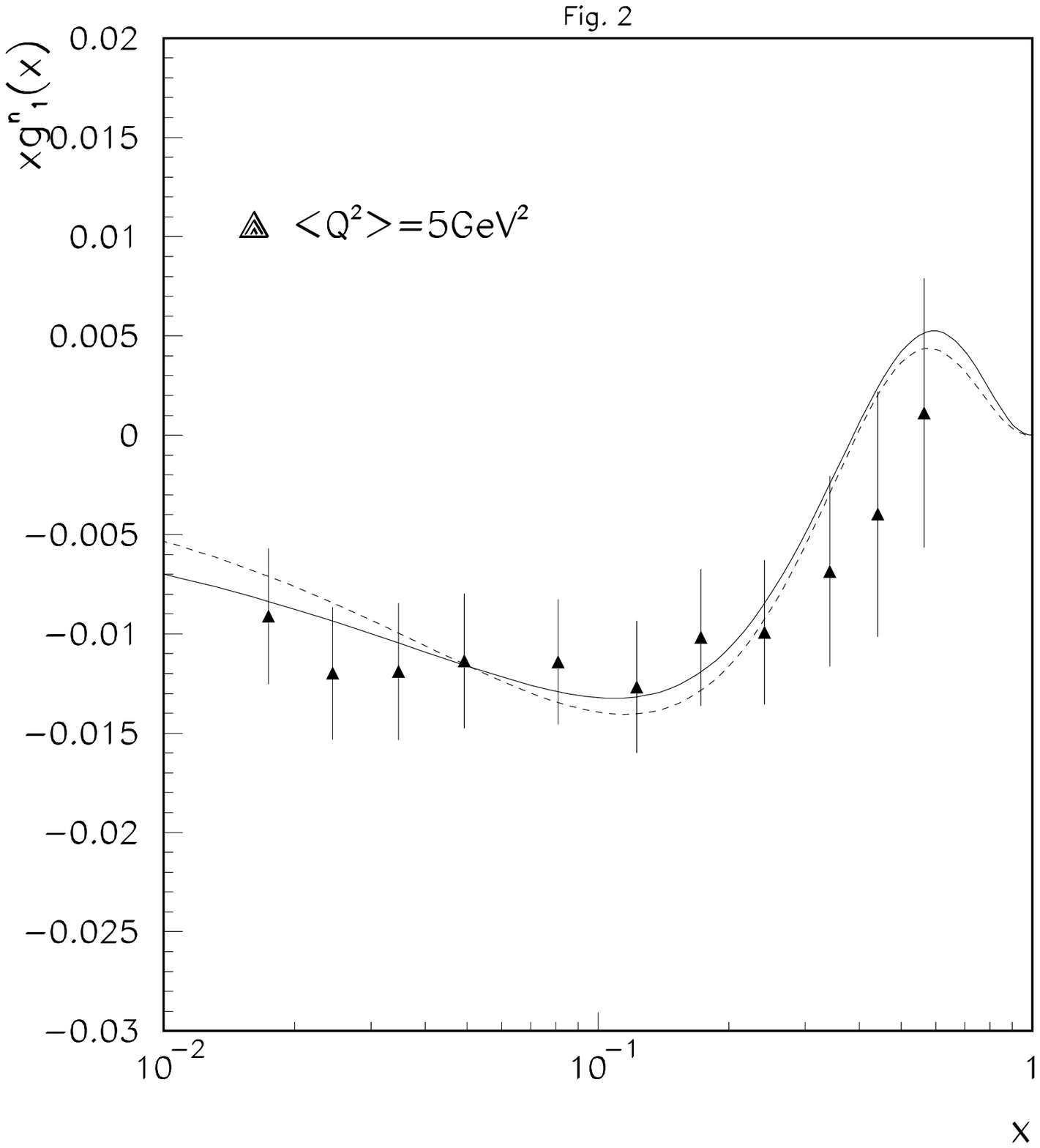,height=15cm}
\caption{The prediction for $x g_1^n(x)$ at $Q^2=3~GeV^2$ in the case $\de
G$ free is plotted and compared with the experimental data
\protect\cite{e154}. The solid and dashed line corresponds to the fit with
present parameterization and with quantum statistical distribution
respectively.} 
\label{f:fig2}
\end{figure}

To get an idea of quality and differences of the two fits we compare in
Fig. \ref{f:fig1} and \ref{f:fig2} $x~F_3(x)$ measured in Ref.~\cite{xf3}
and $x~g_1^n(x)$ measured in Ref.~\cite{e154} with the quantum statistical
and present parameterization in the case $\de G$ free. As we see, the data
are very well reproduced and there is very small difference between the two
fits. 

In conclusion, as a consequence of our analysis, we can add a new fact to
the ones mentioned at the beginning of this paper, supporting a role of the
Pauli principle: $d^\upa$, the valence quark with the lowest first moment,
is also the one with the narrower shape, and the dependence of the shape on
the first moment for each parton is in good agreement with the one found
using Fermi-Dirac functions. 

Our preliminary work on NLO analysis of the SLAC data confirms the trend 
with a slightly narrower distribution for $d^\upa$ and, of course, a 
largely broader one for $u^\upa$ with respect to $u^\dwa$ and $d^\dwa$.

The study of the $Q^2$ evolution is in our plans, especially since the
large $Q^2$ and $x$ excess found at Hera \cite{heraexc} might be explained
by a modification of the evolution, which is resonable if Pauli principle
plays a role in parton distributions, consisting in a slower narrowing of
the distribution of a fermionic parton with its levels almost completely
occupied (in the case of the proton the $u^\upa$), as advocated in
Ref.~\cite{gmcube}.

\newpage

\begin{table}[ht]
\begin{center}
TABLE \ref{t:param} \\
\vspace{.6truecm}
\begin{small}
\begin{tabular}{|c|c|c||c|c|c|}
\hline\hline
\ru1 & \multicolumn{2}{c||}{PP} && \multicolumn{2}{c|}{ST} \\
\hline
\ru1 & {\bf$\de G=0$} & {{\bf $\de G$} free} && {\bf$\de G=0$} & {{\bf$\de
	G$} free} \\
\hline\hline
\ru1 $\al$	& $-0.227\pm0.012$	& $-0.235\pm0.012$
	& $\bea{c}
		A \\ \\
		\al \\ \\
		\be
	\ena$ 
	& $\bea{c}
		\\ 
		2.45\pm0.07 \\ \\
		-0.252\pm0.009 \\ \\
		2.19\pm0.04 \\ \\
	\ena$
	& $\bea{c}
		\\
		2.50\pm0.07 \\ \\
		-0.251\pm0.005 \\ \\
		2.20\pm0.04 \\ \\
	\ena$ \\
\hline
\ru1 $\be_{u^\upa}$	&  $2.35\pm0.04$ & $2.35\pm0.04$
	& $\tilde{x}(u^\upa)$	& 1		& 1 \\
\hline
\ru1 $\be_{d^\dwa}$	& $4.21\pm0.10$	& $4.12\pm0.11$
	& $\tilde{x}(d^\dwa)$	& $0.206\pm0.006$ & $0.182\pm0.011$ \\
\hline
\ru1 $\be_{u^\dwa}$	& $4.35\pm0.16$ & $4.26\pm0.17$
	& $\tilde{x}(u^\dwa)$	& $0.134\pm0.009$ & $0.112\pm0.013$ \\
\hline
\ru1 $\be_{d^\upa}$	& $4.81\pm0.24$ & $4.89\pm0.25$
	& $\tilde{x}(d^\upa)$	& $-0.072\pm0.012$ & $-0.073\pm0.013$ \\
\hline
\ru1 $\be_{\bar{d}}$	& $4.81\pm0.10$ & $4.90\pm0.10$
	& $\tilde{x}(\bar d)$	& $-0.351\pm0.016$ & $-0.388\pm0.024$ \\
\hline
\ru1 $\be_{\bar{u}}$	& $16.4_{-1.8}^{+2.1}$ & $16.8_{-1.8}^{+2.1}$
	& $\tilde{x}(\bar u)$	& $-0.647\pm0.031$ & $-0.715\pm0.046$ \\
\hline
\ru1 $\al_{\de G}$	& --	& $-0.80\pm0.61$
	& $\bea{c} \\
		\tilde{x}(G^\upa) \\ \\
		\tilde{x}(G^\dwa) \\ \\
	\ena$
	& $\bea{c} \\
		-0.129\pm0.009 \\ \\
		-0.129\pm0.009 \\ \\
	\ena$
	& $\bea{c} \\
		-0.119\pm0.010	\\ \\
		-0.191\pm0.029 \\ \\
	\ena$ \\
\hline
\ru1 $\be_{\de G}$		& --		  & $20\pm11$
	& $\overline{x}$	& $0.230\pm0.006$ & $0.243\pm0.006$ \\
\hline
\ru1 $\chi^2$		& 1.89	& 1.90
	& $\chi^2$	& 2.42  & 2.37 \\
\hline
\ru1 Bj			& $0.163\pm0.003$ & $0.162\pm0.003$
	& Bj		& $0.165\pm0.010$ & $0.164\pm0.009$ \\
\hline
\ru1 Gott		& $0.223\pm0.006$ & $0.223\pm0.006$
	& Gott		& $0.209\pm0.012$ & $0.209\pm0.015$ \\
\hline
\end{tabular}
\end{small}
\end{center}
\caption{The values of the parameters found with the present
parameterization (PP) are compared with the ones obtained using quantum
statistical distributions (ST). The {\it liquid} part is given by
Eq.~(\protect\ref{e:liq}). In the last two rows we report the results for
the rhs of the Bjorken \protect\cite{bjork} and Gottfried
\protect\cite{gottfr} sum rules.} 
\label{t:param}
\end{table}

\newpage

\begin{table}[ht]
\begin{center}
TABLE \ref{t:mom1a} \\
\vspace{.6truecm}
\begin{tabular}{|l|c|c|c|c|c|} \hline\hline
\multicolumn{6}{|c|}{\ru1 $\de G = 0$} \\
\hline
\multicolumn{6}{|c|}{\ru1 PP} \\
\hline
\ru1 & $p$ & $p^{(2)}$ & $p^{(2)}/p$ & $p^{(3)}$ & $p^{(3)}/p^{(2)}$ \\
\hline\hline
\ru1 $u^\upa$	& $1.246\pm0.010$ & $0.234\pm0.004$ & $0.187\pm0.003$ & 
$0.081\pm0.002$ & $0.346\pm0.010$ \\
\hline
\ru1 $d^\dwa$	& $0.664\pm0.010$ & $0.086\pm0.002$ & $0.129\pm0.004$ & 
$0.0218\pm0.0009$ & $0.254\pm0.012$ \\
\hline
\ru1 $u^\dwa$	& $0.593\pm0.009$ & $0.075\pm0.002$ & $0.126\pm0.005$ & 
$0.0186\pm0.0010$ & $0.25\pm0.02$ \\
\hline
\ru1 $d^\upa$	& $0.342\pm0.009$ & $0.040\pm0.002$ & $0.117\pm0.006$ & 
$0.0094\pm0.0007$ & $0.23\pm0.02$ \\
\hline
\ru1 $\bar{d}/2$& $0.137\pm0.004$ & $0.0160\pm0.0006$& $0.117\pm0.005$ & 
$0.0037\pm0.0002$ & $0.234\pm0.013$ \\
\hline
\ru1 $\bar{u}/2$& $0.054\pm0.005$ & $0.0023\pm0.0003$& $0.042\pm0.007$ & 
$0.00021\pm0.00005$ & $0.09\pm0.02$ \\
\hline
\multicolumn{6}{|c|}{\ru1 ST} \\
\hline
\ru1 & $p$ & $p^{(2)}$ & $p^{(2)}/p$ & $p^{(3)}$ & $p^{(3)}/p^{(2)}$ \\
\hline\hline
\ru1 $u^\upa$	& $1.25\pm0.05$ & $0.228\pm0.008$ & $0.183\pm0.009$ & 
$0.078\pm0.003$ & $0.34\pm0.02$ \\
\hline
\ru1 $d^\dwa$	& $0.68\pm0.03$ & $0.091\pm0.003$ & $0.134\pm0.007$ & 
$0.0240\pm0.0010$ & $0.26\pm0.02$ \\
\hline
\ru1 $u^\dwa$	& $0.59\pm0.03$ & $0.076\pm0.003$ & $0.128\pm0.008$ & 
$0.0194\pm0.0010$ & $0.26\pm0.02$ \\
\hline
\ru1 $d^\upa$	& $0.35\pm0.02$ & $0.041\pm0.002$ & $0.116\pm0.010$ & 
$0.0097\pm0.0007$ & $0.24\pm0.02$ \\
\hline
\ru1 $\bar{d}/2$& $0.134\pm0.011$ & $0.0144\pm0.0014$ & $0.107\pm0.014$ & 
$0.0033\pm0.0003$ & $0.23\pm0.03$ \\
\hline
\ru1 $\bar{u}/2$& $0.041\pm0.006$ & $0.0043\pm0.0007$ & $0.10\pm0.02$ & 
$0.0010\pm0.0002$ & $0.22\pm0.05$ \\
\hline
\end{tabular}
\end{center}
\caption{The values of the first, second and third moment and of the ratios
$p^{(2)}/p$ and $p^{(3)}/p^{(2)}$ for the {\it gas} component in
Eq.~(\protect\ref{e:distr}) are reported in the case $\de G=0$.} 
\label{t:mom1a}
\end{table}

\begin{table}[ht]
\begin{center}
TABLE \ref{t:mom1b} \\
\vspace{.6truecm}
\begin{tabular}{|l|c|c|c|c|c|} \hline\hline
\multicolumn{6}{|c|}{\ru1 $\de G$ free} \\
\hline
\multicolumn{6}{|c|}{\ru1 PP} \\
\hline
\ru1 & $p$ & $p^{(2)}$ & $p^{(2)}/p$ & $p^{(3)}$ & $p^{(3)}/p^{(2)}$ \\
\hline\hline
\ru1 $u^\upa$	& $1.254\pm0.010$ & $0.233\pm0.004$ & $0.186\pm0.004$ & 
$0.080\pm0.002$ & $0.345\pm0.010$ \\
\hline
\ru1 $d^\dwa$	& $0.661\pm0.011$ & $0.086\pm0.002$ & $0.130\pm0.004$ & 
$0.0220\pm0.0009$ & $0.256\pm0.013$ \\
\hline
\ru1 $u^\dwa$	& $0.592\pm0.009$ & $0.075\pm0.003$ & $0.127\pm0.005$ & 
$0.0189\pm0.0011$ & $0.25\pm0.02$ \\
\hline
\ru1 $d^\upa$	& $0.350\pm0.011$ & $0.040\pm0.002$ & $0.115\pm0.007$ & 
$0.0093\pm0.0007$ & $0.23\pm0.02$ \\
\hline
\ru1 $\bar{d}/2$& $0.138\pm0.004$ & $0.0158\pm0.0006$& $0.115\pm0.005$ & 
$0.0036\pm0.0002$ & $0.230\pm0.013$ \\
\hline
\ru1 $\bar{u}/2$& $0.055\pm0.005$ & $0.0023\pm0.0003$& $0.041\pm0.007$ & 
$0.00021\pm0.00005$ & $0.09\pm0.02$ \\
\hline
\ru1 $\de G$	& $1.9\pm1.3$	  & $0.018\pm0.056$ & $0.009\pm0.030$ & 
$0.0010\pm0.0036$ & $0.05\pm0.26$ \\
\hline
\multicolumn{4}{|c|}{\ru1 ST} \\
\hline
\ru1 & $p$ & $p^{(2)}$ & $p^{(2)}/p$ & $p^{(3)}$ & $p^{(3)}/p^{(2)}$ \\
\hline
\ru1 $u^\upa$	& $1.26\pm0.04$ & $0.229\pm0.008$ & $0.182\pm0.009$ & 
$0.078\pm0.003$ & $0.34\pm0.02$ \\
\hline
\ru1 $d^\dwa$	& $0.66\pm0.02$ & $0.088\pm0.004$ & $0.134\pm0.008$ & 
$0.0235\pm0.0011$ & $0.27\pm0.02$ \\
\hline
\ru1 $u^\dwa$	& $0.57\pm0.02$ & $0.074\pm0.004$ & $0.129\pm0.008$ & 
$0.0193\pm0.0011$ & $0.26\pm0.02$ \\
\hline
\ru1 $d^\upa$	& $0.36\pm0.02$ & $0.044\pm0.003$ & $0.119\pm0.009$ & 
$0.0107\pm0.0007$ & $0.25\pm0.02$ \\
\hline
\ru1 $\bar{d}/2$& $0.131\pm0.013$ & $0.014\pm0.002$ & $0.11\pm0.02$ & 
$0.0034\pm0.0004$ & $0.23\pm0.04$ \\
\hline
\ru1 $\bar{u}/2$& $0.038\pm0.008$ & $0.0040\pm0.0008$ & $0.11\pm0.03$ & 
$0.0009\pm0.0002$ & $0.23\pm0.07$ \\
\hline
\ru1 $\de G$	& $0.91\pm0.04$   & $0.063\pm0.003$ & $0.070\pm0.004$ & 
$0.0113\pm0.0006$ & $0.179\pm0.012$ \\
\hline
\end{tabular}
\end{center}
\caption{Same as table \protect\ref{t:mom1a} with $\de G$ free.}
\label{t:mom1b}
\end{table}

\newpage

\begin{table}[ht]
\begin{center}
TABLE \ref{t:mom2} \\
\vspace{.6truecm}
\begin{small}
\begin{tabular}{|c|c|c|c|c|} \hline\hline
\ru1 & PP & ST & Ref.~\cite{newnlo}  & Ref.~\cite{bupi} \\
\hline\hline
\ru1 $p$	& 0.0908 &  0.0927 & 0.0996	& 0.0989 \\
\hline
\ru1 $p^{(2)}$	& 0.0201 &  0.0197 & 0.0201	& 0.0226 \\
\hline
\ru1 $p^{(2)}/p$& 0.221  &  0.213  & 0.202	& 0.228	\\
\hline
\end{tabular}
\end{small}
\end{center}
\caption{The values of the first and second moment of the non-singlet part
of $g_1^p$ with $\de G$ free, obtained with the present parameterization
(PP), are compared with the same quantities from quantum statistical
distributions (ST) and from Ref.~\protect\cite{newnlo,bupi}. The values
from Ref.~\protect\cite{newnlo} have been reported to $Q^2=3\,GeV^2$ with
the NLO evolution equation for the moments.} 
\label{t:mom2}
\end{table}

\newpage

\newpage

\end{document}